\documentclass{article}
\usepackage{amsmath}
\usepackage{amssymb}
\usepackage{graphicx}
\begin{document}
\begin{center}
\LARGE
\textbf{Non-symmetric transition probability}
\newline
\textbf{in generalized qubit models}
\vspace{0,3 cm}
\normalsize

Gerd Niestegge 
\footnotesize
\vspace{0,2 cm}

Ahaus, Germany

gerd.niestegge@web.de, https://orcid.org/0000-0002-3405-9356
\end{center}
\normalsize
\begin{abstract}
The quantum mechanical transition probability is symmetric.
A probabilistically motivated and more general quantum logical
definition of the transition probability was introduced in two preceding papers without postulating its symmetry, but in all the examples considered there it remains symmetric.
Here we present a class of binary models
where the transition probability 
is not symmetric, using the extreme points of the unit interval
in an order unit space as quantum logic. 
We show that their state spaces 
are strictly convex smooth compact convex sets
and that each such set $K$ gives rise to a 
quantum logic of this class with the state space $K$.
The transition probabilities are symmetric
iff $K$ is the unit ball in a Hilbert space.
In this case, the quantum logic becomes identical
with the projection lattice in a spin factor
which is a special type of formally real Jordan algebra.
\vspace{0,5 cm}

\end{abstract}

\section{Introduction}

A typical feature of common quantum mechanics is 
the symmetry of the transition probability. 
All physical experiments and observations are in line with it,
and attempts to find an axiomatic access to quantum theory 
often postulate it a priori
\cite{araki1980characterization, nie2020charJordan}.
It is also a part of Alfsen and Shultz's so-called 
\emph{pure state properties} \cite{AS02}. 
However, doubts that there is any reasonable physical reason for this postulate 
are as old as the early axiomatic attempts
\cite{guz_1980, guz1980non, mielnik1968geometry, mielnik1969theory}. 
A probabilistically motivated and more general definition 
of the transition probability in the quantum logical framework
was introduced in two preceding papers \cite{nie2020alg_origin, nie2021generic}
without postulating the symmetry, but in all the examples
considered there the transition probability remains symmetric.

A crucial question thus becomes 
whether there are any quantum logical structures 
with a non-symmetric transition probability
and how this probability might behave then. 
Here we identify such structures
and explore some explicit examples.
Models from convexity theory are reused,
which have already been studied 
by Alfsen and Shultz \cite{AS02}
and Berdikulov \cite{berdikulov1990homogeneous, berdikulov1995generalized, 
berdikulov2006notion, berdikulov2010banach}, 
but with a different focus; 
these authors have not explicitly examined the transition probability
and we do this here. Moreover,
using the extreme points of the unit interval
in an order unit space, 
the approach presented here provides a simpler and directer access to
the quantum logics and the transition probabilities than Alfsen and Shultz's theory.
We shall see that our transition probabilities become identical with those
introduced by Mielnik in a different way \cite{mielnik1968geometry, mielnik1969theory}. 

With one exception, the quantum logics which we consider are binary. 
They represent the classical bit, the quantum bit (qubit) and generalized versions 
thereof. 
We use the transition probabilities
to define the new type of binary quantum logics. 
Our main result then becomes an interesting one-to-one relationship 
between the quantum logics of this type and 
the strictly convex smooth compact convex sets.
The characteristics of the quantum logic are 
determined by those of the convex set.
Only if this set is the unit ball in a (pre-)Hilbert space,
the transition probabilities become symmetric.
In this case the quantum logic is the projection lattice in a spin factor.
The spin factors are formally real Jordan algebras \cite{hanche1984jordan}
and are known from the canonical anti-commutator relations for the fermions.
A certain generalization of these spin factors 
will be studied first here and will later motivate 
the definition of the new type of binary quantum logic.

The paper is organized as follows. Sections 2 and 3 briefly sketch
those definitions from the preceding papers \cite{nie2020alg_origin, nie2021generic} 
that will be needed subsequently. 
The generalization of the spin factors
is then introduced in section 4.
The quantum logics, state spaces and transition probabilities
of these generalized spin factors 
are studied in sections 5, 6 and~7.
After a brief intermezzo on the spectral decomposition in section 8,
we come in section 9 to our main results.
The transition probabilities are used to 
characterize the generalized spin factors in a new way and
to define the new type of binary quantum logic. Moreover, 
the relationship to the strictly convex and smooth compact convex sets
is elaborated.
In section~10 we turn to the special generalized spin factors
arising from the spaces $l^{p}$ and $L^{p}$ and 
examine explicit numerical examples with non-symmetric transition probabilities.
Further examples are discussed in section~11.

Some basic knowledge of convex analysis
(extreme points, smoothness, strict convexity, order unit spaces \cite{AS01, AS02})
and the theory of normed linear spaces 
\cite{chidume2009, koethe1969topological}
will be helpful for the reader.

\section{Quantum logics and states}

The quantum logical framework 
that was used in the preceding papers \cite{nie2020alg_origin, nie2021generic} 
shall be recapitulated here first. 

A \textit{quantum logic} is an \emph{orthomodular partially ordered set} $L$ 
with order relation $\leq$, 
smallest element $0$, largest element $\mathbb{I}$ and an ortho\-complement\-ation~$'$.
This means that the following conditions are satisfied by the $e,f \in L$:
\begin{enumerate}
\item[(a)] $ e \leq f$ \textit{implies} $f' \leq e'$.
\item[(b)] $(e')' = e$.
\item[(c)] $e \leq f'$ \textit{implies that} $e \vee f$, \textit{the supremum of} $e$ \textit{and} $f$, \textit{exists}.
\item[(d)] Orthomodular law: $f \leq e$ \textit{implies} $e = f \vee (e \wedge f')$.
\end{enumerate}
Here, $e \wedge f$ denotes the infimum of $e$ and $f$, 
which exists iff $e' \vee f'$ exists. 
Condition (d) is the \emph{orthomodular law}; it implies that
$e \vee e' = \mathbb{I}$ for any $e \in L$.

An element $e \in L$ with $e \neq 0$ is called \emph{minimal} or an \emph{atom}
if there is no $f \in L$ with $f \leq e$ and $0 \neq f \neq e$.
Two elements $e$ and $f$ in $L$ are \emph{orthogonal}, 
if $e \leq f'$ or, equivalently, $f \leq e'$;
in this case, $e \vee f$ exists and shall be noted by $e + f$ in the following.
In a lattice, 
$e \wedge f$ and $e \vee f$ would exist for any elements $e$ and $f$.

A \textit{state} $\mu$ shall allocate probabilities to the 
elements of the quantum logic. Therefore it becomes a map
from $L$ to the unit interval $\left[0,1\right] \subseteq \mathbb{R}$
with $\mu(\mathbb{I})=1$ and $\mu(e + f) = \mu(e) + \mu(f)$ for any two
orthogonal elements $e$ and $f$ in $L$.
A set $S$ of states on $L$ is called 
\textit{strong} if, for any $e,f \in L$,
$$ \left\{ \mu \in S \:|\: \mu(e) = 1 \right\} \subseteq 
\left\{ \mu \in S \:|\: \mu(f) = 1 \right\} \ \  \Rightarrow \ \  e \leq f .$$
Note that this implies that there is a state $\mu \in S$ with $\mu(e) = 1$ 
for each $e \in L$ with $e \neq 0$. 

\section{Transition probability in quantum logics}

We can now restate the following
definition of the transition probability from
the preceding papers \cite{nie2020alg_origin, nie2021generic}. 
\vspace{0,3 cm}

\textbf{Definition 3.1} 
\itshape 
Let $L$ be a quantum logic and $S$ a strong set of states on $L$.
If a pair $e,f \in L$ with $e \neq 0$ and some $s \in [0,1]$
satisfies the condition 
\begin{center}
$\mu(f) = s$ for all $\mu \in S$ with $\mu(e)=1$,
\end{center}
then $s$ is called the \emph{transition probability from} $e$ \emph{to} $f$ 
and is denoted by $\mathbb{P}(f|e)$.
\normalfont
\vspace{0,3 cm}

\noindent
The identity
$\mathbb{P}(f|e) = s$ is equivalent to the set inclusion
\begin{equation*}
\left\{\mu \in S \:|\: \mu(e)=1\right\} \subseteq \left\{\mu \in S \:|\: \mu(f)=s\right\}
\end{equation*}
and means that, whenever the probability of $e$ is $1$, 
the probability of $f$ is fixed and its numerical value is $s$;
particularly in the situation after a quantum measurement
that has provided the outcome $e$, the probability of $f$ 
becomes $s$, independently of any initial state 
before the measurement.

Two elements $e$ and $f$ in $L$ 
are orthogonal iff $\mathbb{P}(f | e) = 0$
and $e \leq f$ holds iff $\mathbb{P}(f | e) = 1$.
The second part here holds since $S$ is a strong set of states,
and the first part follows by considering $f'$.

In a preceding papers \cite{nie2020alg_origin}
it was elaborated that,
in the case of a Hilbert space quantum logic
with its usual state space,
this transition probability becomes identical with the usual 
quantum mechanical transition probability:
$$ \mathbb{P}(f | e) = \left|\left\langle \varphi |\psi \right\rangle\right|^{2}, $$
where $\varphi$ and $\psi$ are normalized vectors in the Hilbert space
and the atoms $e$ and $f$ are the one-dimensional subspaces generated by $\varphi$ and $\psi$
(or the corresponding self-adjoint projection operators). Therefore the \emph{symmetry condition}
$$\mathbb{P}(f | e) = \mathbb{P}(e | f)$$
holds. The quantum logics formed by the projection lattices in the
von Neumann algebras and formally real Jordan algebras
(with their usual state spaces)
satisfy the same symmetry condition~\cite{nie2021generic}.
The following lemma demonstrates an interesting consequence of this symmetry.
\vspace{0,3 cm}

\textbf{Lemma 3.2} 
\itshape 
Let $L$ be a quantum logic and $S$ a strong set of states on $L$. 
Moreover, suppose that the transition probability 
$\mathbb{P}(f | e)$ exists for every atom $e$ and every element $f$ in $L$
and that $\mathbb{P}(f | e) = \mathbb{P}(e | f)$ holds for any two atoms
$e$ and $f$. If $e_1 + e_2 +... + e_m = f_1 + f_2 + ... f_n$ with two families
of atoms $e_1, e_2, ..., e_m$ and $f_1, f_2, ..., f_n$ such that
the atoms in the same family are pairwise orthogonal, then $m = n$ must hold. 
\normalfont
\vspace{0,3 cm}

Proof. Under the above assumptions we have 
$f_k \leq \sum_l e_l$ and $\mathbb{P}(\sum_l e_l|f_k) = 1$
for each $k$ as well as $e_l \leq \sum_k f_k$ and
$\mathbb{P}(\sum_k f_k|e_l) = 1$ for each $l$. Then 

\vspace{0,3 cm}
\hspace*{10mm}
$ \sum_k \sum_l \mathbb{P}(e_l|f_k) 
= \sum_k \mathbb{P}(\sum_l e_l|f_k) = n$ 
and 

\vspace{0,3 cm}
\hspace*{10mm}
$ \sum_k \sum_l \mathbb{P}(e_l|f_k) 
= \sum_l \sum_k \mathbb{P}(f_k|e_l) 
= \sum_l \mathbb{P}(\sum_k f_k|e_l) 
= m$. \hfill $\square$
\vspace{0,3 cm}

Lemma 3.2 means that a kind of dimension function exists, 
if the transition probabilities are symmetric.

The existence of the transition probability $\mathbb{P}(f|e)$
for two elements $e$ and $f$ in a quantum logic $L$
does not require that $e$ is an atom \cite{nie2020alg_origin, nie2021generic}.
However, if $\mathbb{P}(f|e)$ exists for \emph{all} $f \in L$,
$e$ must be an atom \cite{nie2021generic}
and there is only one unique
state $\mu_e$ with $\mu_e(e) = 1$; $\mu_e(f) = \mathbb{P}(f|e)$
for $f \in L$. If there is only one unique state $\mu_e$
with $\mu_e(e) = 1$ for some $e \in L$,
$\mathbb{P}(f|e)$ exists for all $f \in L$ with 
$\mathbb{P}(f|e) = \mu_e(f)$ and $e$ must be an atom.
In this case, $\mu_e$ is an 
extreme point in the state space $S$
[$\mu_e = t \nu_1 + (1-t) \nu_2$ with 
$\nu_1,\nu_2 \in S$ and $0 < t < 1$
implies $ 1 = \mu_e(e) = t \nu_1(e) + (1-t) \nu_2(e)$,
thus $\nu_1(e) = 1 = \nu_2(e)$ and $\nu_1 = \nu_2 = \mu_e$].
The extreme points in the state space $S$
are usually called \emph{pure} states.

\section{Definition of the generalized spin factors}

A special class of quantum logics arises from the 
so-called generalized spin factors,
which were studied by Berdikulov 
\cite{berdikulov1990homogeneous, berdikulov1995generalized, 
berdikulov2006notion, berdikulov2010banach}.
The following notions from the theory of normed linear spaces
will be needed to define them.

A normed linear space $X$ is called \emph{smooth} if 
its unit ball is smooth. This means that,
for every $x \in X$, $\left\|x\right\| = 1$,
there exists a \emph{unique} bounded linear functional $\rho_x$ in its dual 
such that $\left\| \rho_x \right\| = 1$ and $\rho_x (x) = 1$. 
A normed linear space $X$ is called \emph{strictly convex}, if
its unit ball is strictly convex. This means that
we have $ \left\| t x +(1-t) y \right\| < 1$ 
for all $x,y \in X$, $x\neq y$, $\left\|x\right\| = 1 = \left\|y\right\|$
and $t \in \mathbb{R}$ with $0 < t < 1$ 
or, equivalently, that each $x \in X$ with $\left\|x\right\| = 1$ 
is an extreme point of the closed unit ball \cite{chidume2009}.

Examples of normed linear spaces that are both smooth 
and strictly convex are the Hilbert spaces,
the $L^{p}$ and $l^{p}$ spaces
with the norm $\left\| \ \right\|_p$, $1 < p < \infty$ 
\cite{chidume2009, koethe1969topological}. 
Concerning their mathematical structure,
the smooth and strictly convex
normed spaces are closer to the Hilbert spaces 
than the general normed spaces.

Now let $X$ be any normed $\mathbb{R}$-linear space
and consider the direct sum $A := X \oplus \mathbb{R}$
with the following order relation:
$0 \leq x \oplus s $ 
iff $\left\|x\right\| \leq s$ 
for $x \in X$, $s \in \mathbb{R}$.
Moroever, define $\mathbb{I} := 0 \oplus 1$. 
It is not hard to verify that 
$A$ becomes an order unit space with order unit $\mathbb{I}$
in this way and that $A$ has the following norm: 
$\left\|x \oplus s\right\| = \left\|x\right\| + |s|$ 
for $x \in X$ and $s \in \mathbb{R}$.

In the case of a Hilbert space $X$ over the real numbers $\mathbb{R}$
with the inner product $\left\langle \ | \ \right\rangle$, 
$A$ becomes a formally real Jordan algebra \cite{AS02, hanche1984jordan}, 
when it is equipped with the following product:
$ (x \oplus s) \circ (y \oplus t) := (tx + sy) \oplus (\left\langle x|y\right\rangle + st)$ 
for $x,y \in X$ and $s,t \in \mathbb{R}$.
This special type of Jordan algebra is usually 
called \emph{spin factors} \cite{AS02, hanche1984jordan}.
They represent the canonical anti-commutator relations for the fermions, 
since we have $(x_k \oplus 1) \circ (x_l \oplus 1) = \delta_{kl} \mathbb{I} $
for any orthonormal basis $x_k$ in the Hilbert space $H$
and since the Jordan product $\circ$ becomes identical with the 
anti-commutator $\left[ \ , \ \right]^{+}$,
when $A$ is represented as operator algebra
(which is generally possible for the spin factors \cite{AS02, hanche1984jordan}).

Therefore Berdikulov chose the name \emph{generalized spin factor}
for the order unit spaces $A = X \oplus \mathbb{R}$,
arising from the \emph{smooth and strictly convex} normed linear spaces $X$,
but required that $X$ is the dual of some other space
\cite{berdikulov1990homogeneous, berdikulov1995generalized, berdikulov2006notion, berdikulov2010banach}. 
Note that, in this paper, we do neither require that $X$ is complete nor that $X$ 
is a dual space, and we call $A = X \oplus \mathbb{R}$ 
a \emph{generalized spin factor} (\emph{spin factor}) also 
in those cases when $X$ is just a \emph{smooth and strictly convex
normed $\mathbb{R}$-linear space} (\emph{pre-Hilbert space}).
Moreover we include the case $X = \mathbb{R}$ which is usually ruled out,
because $A = X \oplus \mathbb{R}$ is decomposable then
and does not satisfy the criteria of a factor.

\section{The quantum logic of a generalized spin factor}

Instead of using Alfsen and Shultz's compressions, projective faces or
projective units \cite {AS02} of the state space, 
we provide a rather direct and simple access
to the quantum logics of the generalized spin factors.
Consider the unit interval $[0,\mathbb{I}]$ 
in $A = X \oplus \mathbb{R}$
with a smooth and strictly convex normed $\mathbb{R}$-linear space $X$.
\begin{align}
[0,\mathbb{I}] :&= \left\{ a \in A : 0 \leq a \leq \mathbb{I} \right\} 
= \left\{ a \in A : 0 \leq a \text{ and } \left\|a\right\| \leq 1 \right\} \\
&= \left\{ x \oplus s : x \in X, s \in \mathbb{R}, \left\| x \right\| \leq s
\text{ and } \left\| x \right\| \leq 1 - s  \right\} \\
&= \left\{ x \oplus s : x \in X, s \in \mathbb{R}, \left\|x\right\| \leq s \leq 1 - \left\|x\right\| \right\}\\
&\subseteq \left\{ x \oplus s : x \in X, \left\|x\right\| \leq \frac{1}{2} \text{ and } 0 \leq s \leq 1 \right\}.
\end{align}
The projection lattice in a von Neumann algebra or formally real Jordan algebra
and particularly in a spin factor
is identical with the extreme points of the the positive part of the unit ball 
\cite{AS01} and therefore
the extreme points of $[0,\mathbb{I}]$ shall become our quantum logic~$L_X$:
$$L_X := ext([0,\mathbb{I}]).$$
The order relation is inherited from the one on $A$. 
The orthocomplement of $e \in L_X$ is $e' := \mathbb{I} - e$.
Note that $e$ is extremal iff $\mathbb{I} - e$ is extremal in $[0,\mathbb{I}]$.
\vspace{0,3 cm}

\textbf{Proposition 5.1} 
$L_X = ext([0,\mathbb{I}]) = \left\{ 0, \mathbb{I} \right\} \cup \left\{ \frac{1}{2}(x \oplus 1) : x \in X, \left\|x\right\| = 1 \right\} $.
\vspace{0,3 cm}

Proof. In any order unit space, $0$ and $\mathbb{I}$ are extreme points of $[0,\mathbb{I}]$.
Now suppose $x \in X$, $\left\|x\right\| = 1$ and 
\begin{align*}
\frac{1}{2}(x \oplus 1) 
&= t (y_1 \oplus s_1) + (1-t) (y_2 \oplus s_2) \\
& = (t y_1 + (1-t) y_2) \oplus (t s_1 + (1-t) s_2) 
\end{align*}
with $y_1 \oplus s_1, y_2 \oplus s_2 \in [0,\mathbb{I}]$ and $0 < t < 1$.
Then $\frac{1}{2} x = t y_1 + (1-t) y_2$ and $ \frac{1}{2} = t s_1 + (1-t) s_2 $.
Since $0 \leq \left\|y_1\right\|,\left\|y_2\right\| \leq \frac{1}{2}$ by (4),
the strict convexity implies $y_1 = y_2 = \frac{1}{2}x$.
Since $\left\|y_1\right\| =\left\| y_2\right\| = \frac{1}{2}$ 
and $y_1 \oplus s_1, y_2 \oplus s_2 \in [0,\mathbb{I}]$,
we get by (3) $s_1 = s_2 = \frac{1}{2}$. Therefore
$\frac{1}{2}(x \oplus 1)$ is an extreme point in $[0,\mathbb{I}]$.

Now let $x \oplus s$ be any element in $[0,\mathbb{I}]$ with $x \neq 0$
and $\left\|x\right\| < \frac{1}{2}$. It becomes 
a convex combination of the three elements $0$, $\mathbb{I}$ and 
$\frac{1}{2}(\frac{x}{\left\|x\right\|} \oplus 1)$ in $[0,\mathbb{I}]$
with the parameters $1 - s - \left\|x\right\|$, $s - \left\|x\right\|$ 
and $2 \left\|x\right\|$. These three parameters are non-negative real numbers 
and their sum becomes $1$. Since $0 < 2 \left\|x\right\| < 1$, 
$x \oplus s$ is not an extreme point in $[0,\mathbb{I}]$.
Moreover, $0 \oplus s$ is not an extreme point unless $s = 0$ or $s = 1$, 
since in the other cases we have $0 \oplus s = (1-s) 0 + s \mathbb{I}$.
In the remaining cases we have $\left\|x\right\| = \frac{1}{2}$; 
then $s = \frac{1}{2}$ by (3) and $x \oplus s$ becomes an extreme point as 
shown above.\hfill $\square$
\vspace{0,3 cm}

Note that only the strict convexity is used in the proof of 
Proposition~5.1; the smoothness of $X$ is not needed here.

With $x \in X$, $\left\|x\right\| = 1$, the orthogonal complement 
of $\frac{1}{2}(x \oplus 1) \in L_X$ 
is $\frac{1}{2}(-x \oplus 1) \in L_X$.
All these elements are atoms in the quantum logic $L_X$;
only $0$ and $\mathbb{I}$ are not atoms. 
The set of atoms in $L_X$ 
thus becomes isomorphic to the unit sphere in $X$. 

The quantum logic $L_X$ is a lattice, since the supremum $e \vee f$ 
and the infimum $e \wedge f$ exist for all $e,f \in L_X$:
$0 \vee f = f$, $\mathbb{I} \vee f = \mathbb{I}$, 
$0 \wedge f = 0$, $\mathbb{I} \wedge f = f$,
$f \vee f = f = f \wedge f $ for all $f \in L_X$, and
$e \vee f = \mathbb{I}$, $e \wedge f = 0$ for any two atoms $e$ and $f$ with $e \neq f$.
The orthomodular law is satisfied since 
$e \leq f$ is possible for $e,f \in L_X$ only if either
$e = 0$ or $f = \mathbb{I}$ or $e = f$.

Since the maximum number of pairwise orthogonal atoms is two, 
each quantum logic $L_X$ with any 
smooth and strictly convex normed $\mathbb{R}$-linear space
$X$ represents a binary model.
In the cases $X = \mathbb{R}^{n}$ with the Euclidean norm
we get the quantum logic 
of the usual complex qubit with $n = 3$ (Bloch sphere),
the real version of the qubit with $n = 2$ 
and the classical bit with $n=1$. 
\newpage 

\section{The state space of a generalized spin factor}

The state space $S$ of an order unit space $A$ consists 
of those $\mathbb{R}$-linear functionals $\mu : A \rightarrow \mathbb{R}$
that satisfy $\mu(\mathbb{I}) = 1$ and 
$\mu(a) \geq 0$ for all $a \in A$ with $a \geq 0$.
For the quantum logic $L_X$, defined in the previous section,
we consider the state space $S_X$ that consists of the 
restrictions of the states on $A = X \oplus \mathbb{R}$ to $L_X$.
The restriction of a state on $A$ to $X$ becomes a bounded linear 
functional $\rho$ on $X$ with $\left\|\rho\right\| \leq 1$. 
Vice versa, each bounded linear functional $\rho$ on $X$ 
with $\left\|\rho\right\| \leq 1$ extends to a unique state $\mu$ on $A$ by
$\mu(x \oplus s) := \rho(x) + s$. 
The state space of $A$ and $L_X$ 
thus becomes isomorphic to the unit ball of the dual of $X$.

The state resulting from
$\rho = 0$ is called the \emph{trace state} and is denoted by $\mu_{tr}$ here.
It allocates the same number ($1/2$) to each atom 
and $2 \mu_{tr}$ becomes a kind of a dimension function.
\vspace{0,3 cm}

\textbf{Lemma 6.1} 
(i) 
\itshape For each atom $e$ in $L_X$ there is one unique state $\mu_e$ in $S_X$
with $\mu_e(e) = 1$. With $e = \frac{1}{2}(x \oplus 1)$, $x \in X$ and $\left\|x\right\| = 1 $,
the restriction of $\mu_e$ to $X$ is identical with the unique $\rho_x$ 
which results from the smoothness of $X$ 
and which satisfies $\left\|\rho_x\right\| =1$ and $\rho_x(x)= 1$.
\normalfont

(ii)
\itshape The state space $S_X$ of the quantum logic $L_X$ is strong.
\normalfont
\vspace{0,3 cm}

Proof. (i) Suppose $e = \frac{1}{2}(x \oplus 1)$ with $x \in X$ and $\left\|x\right\| = 1 $.
Then $\mu(e) = 1$ with $\mu \in S_A$ iff the restriction of $\mu$ to $X$
allocates $1$ to $x$. Due to the smoothness of $X$, there is one unique
such functional $\rho_x$. 

(ii) To show that $S_X$ is strong, it is sufficient to consider 
$e,f \in L_X $, both different from $0$ and $\mathbb{I}$, with
$\left\{ \mu \in S_X \:|\: \mu(e) = 1 \right\} \subseteq 
\left\{ \mu \in S_X \:|\: \mu(f) = 1 \right\}$. 
However, by (i) there is only one state $\mu_e$
with $\mu_e (e) = 1$ and only one state
$\mu_f$ with $\mu_f (f) = 1$. Therefore  $\mu_e = \mu_f$.
From $e = \frac{1}{2}(x \oplus 1)$ and $f = \frac{1}{2}(y \oplus 1)$ 
with $x,y \in X$ and $\left\|x\right\| = 1 = \left\|y\right\|$ 
we get by (i) $\rho_x = \rho_y =: \rho$;
this means $\rho(x) = 1 = \rho(y)$.
For all $t \in [0,1]$ then
$1 = \rho(tx + (1-t)y) \leq \left\|tx + (1-t)y\right\| \leq t\left\|x\right\| + (1-t)\left\|y\right\| =1$
and therefore $\left\|tx + (1-t)y\right\| =1$.
From the strict convexity of $X$ we get $ x = y$ and $e = f$, particularly $e \leq f$.
\hfill $\square$

\section{Transition probability in a generalized spin factor}

By Lemma 6.1 (i), the transition probability $\mathbb{P}(f|e)$ exists 
for each atom $e$ and all elements $f$ in the quantum logic $L_X$,
arising from any smooth and strictly convex normed space $X$ as
explained in the previous three sections.
Moreover, using the same notation as in Lemma 6.1,
with $e = \frac{1}{2}(x + \mathbb{I})$,
$f = \frac{1}{2}(y + \mathbb{I})$, $x,y \in X$ 
and $\left\|x\right\| = 1 = \left\|y\right\|$, we have
\begin{equation}
\mathbb{P}(f|e) = \mu_e(f) = \frac{1}{2}(\rho_x(y) + 1).
\end{equation}
Since $e' = \frac{1}{2}(-x + \mathbb{I})$
and $\rho_{-x} = - \rho_x$, this implies the identity
\begin{equation}
\mathbb{P}(f|e) + \mathbb{P}(f|e') = 1.
\end{equation}
In sections 9 and 11, we shall encounter other quantum logics 
where (6) does not hold.
\vspace{0,3 cm}

\textbf{Theorem 7.1}
\itshape
The following two conditions are equivalent:
\normalfont
\begin{enumerate}
\item[(i)]
\itshape
$\mathbb{P}(f|e) = \mathbb{P}(e|f)$ holds for the atoms $e$ 
and $f$ in the quantum logic $L_X$ (this means that the transition 
probability is symmetric).
\normalfont
\item[(ii)]
\itshape
$X$ is a pre-Hilbert space (this means that 
$A = X \oplus \mathbb{R}$ is a spin factor).
\normalfont
\end{enumerate}
\vspace{0,3 cm}

Proof.
(i) $\Rightarrow$ (ii):
Suppose that
$\mathbb{P}(f|e) = \mathbb{P}(e|f)$ for the atoms $e$ 
and $f$ in the quantum logic $L_X$. 
For $x = 0$ or $y = 0$ in $X$ we define $ \left\langle x | y\right\rangle := 0$.

For $x \neq 0 \neq y$ in $X$ we define
$ \left\langle x | y\right\rangle := \rho_x(y)$,
which is linear in the second argument $y$.
Again $\rho_x$ denotes the unique functional on $X$ with 
$\left\|\rho_x\right\| = 1$ and $\rho_x(x) = \left\|x\right\|$,
which exists due to the smoothness of $X$.
Now $e_x := \frac{1}{2}(\frac{x}{\left\|x\right\|} + \mathbb{I})$
and $e_y := \frac{1}{2}(\frac{y}{\left\|y\right\|} + \mathbb{I})$ 
are atoms and 
\begin{align*}
\left\langle x | y\right\rangle = \rho_x(y) &= \left\|x\right\| \ \left\|y\right\| \ \rho_{\frac{x}{\left\|x\right\|}} (\frac{y}{\left\|y\right\|}) \\
&= \left\|x\right\| \ \left\|y\right\| \ ( 2 \mathbb{P}(e_y | e_x) - 1)
\end{align*}
by (5). Then $\left\langle x | x \right\rangle = \left\|x\right\|^{2} > 0$.
The symmetry of the transition probability implies 
$\left\langle x | y\right\rangle = \left\langle y | x\right\rangle$
and the linearity in the first argument.
In this way $X$ becomes a pre-Hilbert space.

(ii) $\Rightarrow$ (i): Use equation (5) and note that
$\rho_x(y) = \left\langle x|y \right\rangle$ and 
$\left\langle x|y \right\rangle = \left\langle y|x \right\rangle$
for the elements $x,y$ in the real pre-Hilbert space $X$
with $\left\|x\right\| = 1 = \left\|y\right\|$. 
\hfill $\square$

\section{Spectral decomposition in a generalized spin factor}

If $ 0 \neq x \in X$ and $s \in \mathbb{R}$, the element $x \oplus s$ 
in the order unit space $A = X \oplus \mathbb{R}$ 
with any smooth and strictly convex normed space $X$ can be written as 
$$x \oplus s = 
(s + \left\|x\right\| ) \: e  
+ (s - \left\|x\right\| ) \: e'$$
with the orthogonal atoms 
\begin{center}
$e := \frac{1}{2}\left(\frac{x}{\left\|x\right\|} \oplus 1\right) $ 
and $e' = \frac{1}{2}\left(\frac{-x}{\left\|x\right\|} \oplus 1\right)$.
\end{center}
This representation can be considered as the spectral decomposition of $x \oplus s$,
and $x \oplus s$ can be interpreted as an observable with the potential
measurement outcomes $s + \left\|x\right\|$ and $s - \left\|x\right\|$.
The spectral decomposition is unique; this will be shown in Theorem 9.1 (i)
under more general assumptions.

Any functions and particularly the power functions 
can then be applied to $x \oplus s$ in the usual way:
\begin{equation}
\left(x \oplus s\right)^{n} := 
(s + \left\|x\right\| )^{n} \: e
+ (s - \left\|x\right\| )^{n} \: e'.
\end{equation}
Then $0 \leq \left(x \oplus s\right)^{2}$, and the quantum logic $L_X$ 
becomes identical with the idempotent elements in $A$.
The product in a Jordan algebra satisfies the identity 
\begin{equation}
a \circ b = \frac{1}{4} \left((a + b)^{2} - (a - b)^{2}\right)
\end{equation}
and therefore the question arises whether this identity can be used 
to derive a product in $A = X \oplus \mathbb{R}$. The bilinearity
of this product again requires that $X$ is a pre-Hilbert space
(and the transition probabilities become symmetric),
since an inner product is then defined on $A$ and 
particularly on the subspace $X$ by
$\left\langle a | b \right\rangle := \mu_{tr}(a \circ b) $ for
$a,b \in A$. Here $\mu_{tr}$ denotes the trace state on $A$ (see section 6)
and by (7) we get
$\mu_{tr}(x ^{2}) = \left\|x\right\|^{2}$ 
for $x \in X$; therefore $ \left\langle x | x \right\rangle = \mu_{tr}(x^{2}) \geq 0$
and $ \left\langle x | x \right\rangle = 0$ iff $x = 0$.

It is interesting to note that 
any generalized spin factor 
with the power function defined by (7) 
satisfies Segal's postulates for a system of observables 
\cite{segal1947postulates}.
Other examples, which 
demonstrate that Segal's postulates are not sufficient 
to get a bilinear product for the observables, 
were presented very early by Lowdenslager \cite{lowdenslager1957postulates}
and Sherman \cite{sherman1956segal}, 
but so far nobody has studied the connection 
with the transition probabilities.
The above considerations together with Theorem 7.1 
and the fact that spin factors are Jordan algebras
show that, in the case of the generalized spin factors, 
the symmetry of the transition probabilities is equivalent 
to the bilinearity of the product (8).

Generally, in Lowdenslager's and Sherman's examples, 
the idempotent elements (their quantum logic) do not coincide with the extreme points
of the unit interval (our quantum logic),
the state space is not strong and
the transition probabilities do not exist for their quantum logic.
This means that these examples are more bizarre than the 
generalized spin factors and further cases 
which we are going to consider in sections  9 and 11.

\section{Axiomatic characterization}

Instead of starting from a smooth and strictly convex normed linear space $X$ 
and constructing the direct sum $X \oplus \mathbb{R}$, we shall now 
identify those characteristics of an order unit space that make it
a generalized spin factor.
We will not use Alfsen and Shultz's theory 
of spectral convex sets \cite{AS02}
as Berdikulov does in his different characterization~\cite{berdikulov2010banach}.
In our approach the transition probabilities will play an important role.
\vspace{0,3 cm}

\textbf{Theorem 9.1} 
\itshape
Let $A$ be an order unit space 
with order unit $\mathbb{I}$. 
As our quantum logic
we consider the
extreme points $L_A := ext[0,\mathbb{I}]$ of the unit interval
and the state space $S_A$ consists of the normalized 
positive $\mathbb{R}$-linear functionals on $A$.
Moreover, we assume that 
\begin{enumerate}
\normalfont
\item[(a)]
\itshape
the state space $S_A$ is strong for the quantum logic $L_A$
with the order relation inherited from $A$ and $e' = \mathbb{I} - e$ for $e \in L_A$,
\normalfont
\item[(b)]
\itshape
$0$ and $\mathbb{I}$ are the only elements in $L_A$ that are not atoms, 
\normalfont
\item[(c)]
\itshape
each $a \in A$ has a spectral decomposition
$a = s e + t e'$ with $s,t \in \mathbb{R}$, an atom $e$ 
and its orthogonal complement $e'$ (which is also an atom),
\normalfont
\item[(d)]
\itshape
the transition probability $\mathbb{P}(f|e)$
exists for each atom $e$ and $f \in L_A$. 
\end{enumerate}
Then:
\normalfont
\begin{enumerate}
\item[(i)] 
\itshape
The spectral decomposition is unique for $a \notin \mathbb{R} \mathbb{I}$.
\normalfont
\item[(ii)]
\itshape 
The state space $S_A$ is a strictly convex and smooth w*-compact convex subset 
of the hyperplane $\left\{\rho \in A^{*} : \rho(\mathbb{I})=1\right\}$ 
in the dual $A^{*}$ of $A$.
Moreover, $\partial S_A = ext(S_A) = \left\{\mu_e: e \text{ is an atom} \right\}$. Here
$\partial S_A$ and $ext(S_A)$ denote the topological boundary 
(in the hyperplane)
and the extreme boundary of $S_A$.
\end{enumerate}
Moreover:
\normalfont
\begin{enumerate}
\item[(iii)]
\itshape
$A$ is a generalized spin factor iff equation \normalfont (6) \itshape holds
for the transition probabilities of any atoms $e$ and $f$.
\normalfont
\item[(iv)]
\itshape
$A$ is a spin factor iff the transition probabilities are symmetric.
\normalfont
\item[(v)]
\itshape
$A$ is a spin factor iff the spectral decomposition results in a bilinear product via 
$a_1 \circ a_2 := \frac{1}{4}((a_1 + a_2)^{2} - (a_1 - a_2)^{2})$ for $a_1, a_2 \in A$ and
$a^{2} := s^{2}e + t^{2}e'$ for $a = se + te'$, $s,t \in \mathbb{R}$ and an atom $e$. 
\end{enumerate}
\normalfont
\vspace{0,3 cm}

Proof. 
For each atom $e$ let $\mu_e \in S_A$
denote the unique state with $\mu_e(e) = 1$. This means 
$\mu_e(f) = \mathbb{P}(f|e)$. First we show that
$0 \leq s e + t e'$ iff $0 \leq s,t$ and
$\left\|s e + t e'\right\| = max\left\{\left|s\right|,\left|t\right|\right\}$ 
for each atom $e$. 

If $0 \leq s e + t e'$, use $\mu_e$ and $\mu_{e'}$
to get $0 \leq s,t$. The only-if part is obvious.
From
\begin{align*}
min\left\{s,t\right\} \mathbb{I} &= min\left\{s,t\right\} e + min\left\{s,t\right\} e'
\leq s e + t e' \\
&\leq max\left\{s,t\right\} e + max\left\{s,t\right\} e' 
= max\left\{s,t\right\} \mathbb{I}
\end{align*}
we get $\left\|s e + t e'\right\| \leq max\left\{\left|s\right|,\left|t\right|\right\}$ 
and then use $\mu_e$ and $\mu_{e'}$ again to get 
$$\left\|s e + t e'\right\| = max\left\{\left|s\right|,\left|t\right|\right\}.$$

(i) Suppose
$a = s_1 e_1 + t_1 {e_1}' = s_2 e_2 + t_2 {e_2}'$
with $s_1,s_2,t_1,t_2 \in \mathbb{R}$ and atoms $e_1,e_2$.
If $s_1 = t_1$, we get $s_1 \mathbb{I} = s_2 e_2 + t_2 {e_2}'$
and with $\mu_{e_2}$ and $\mu_{{e_2}'}$ then $s_1 = s_2 = t_2$
and $a = s_1 \mathbb{I} \in \mathbb{R} \mathbb{I}$.
If $s_1 \neq t_1$, then 
$s_2 e_2 + t_2 {e_2}' = s_1 e_1 + t_1 {e_1}' = (s_1 - t_1) e_1 + t_1 \mathbb{I}$
and $e_1 \in \mathbb{R} e_2 \oplus \mathbb{R} {e_2}'$. Since $e_1$ is extremal
in $[0,\mathbb{I}]$, $e_1$ is also extremal in 
$[0,\mathbb{I}] \cap (\mathbb{R} e_2 \oplus \mathbb{R} {e_2}')$. 
The extreme points of this set are $0, \mathbb{I}, e_2, {e_2}'$ and,
since $e_1$ is an atom, we have 
either 
$e_1 = e_2$ and then ${e_1}' = {e_2}'$, $s_1 = s_2$, $t_1 = t_2$ 
or 
$e_1 = {e_2}'$ and then ${e_1}' = e_2$, $s_1 = t_2$, $t_1 = s_2$.

(ii) $S_A$ is a w*-compact convex subset 
of the hyperplane $\left\{\rho \in A^{*} : \rho(\mathbb{I})=1\right\}$.
Let $\nu_o$ belong to the topological boundary of $S_A$
in this hyperplane.
We shall first show that $\nu_o = \mu_e$ with some atom $e$.
By the Hahn-Banach separation theorem, there is 
an element $\mathbb{I} \neq a \in A$ with 
$\nu_o(a) = 1$ and $\nu(a) \leq 1$ for all $\nu \in S_A$;
$a = s_1 e + s_2 e'$ with an atom $e$, $s_1,s_2 \leq 1$
and $s_1 < 1$ or $s_2 < 1$.
Then $1 = s_1 \nu_o(e) + s_2 \nu_o(e') = s_1 \nu_o(e) + s_2 (1 - \nu_o(e))$
and therefore either $\nu_o(e) = 0$, $\nu_o(e') = 1$ and $\nu_o = \mu_{e'}$
or $\nu_o(e) = 1$ and $\nu_o = \mu_{e}$.
Thus we have $\partial S_A \subseteq \left\{\mu_e: e \text{ is an atom} \right\}$.

Since $\left\{\mu_e: e \text{ is an atom} \right\} 
\subseteq ext(S_A) \subseteq \partial S_A$ 
generally holds, these three sets coincide. 
Particularly each topological boundary point is
an extreme point and therefore $S_A$ is strictly convex
in the hyperplane.

We now prove that $S_A$ is smooth in the hyperplane
at the point $\mu_f$ for each atom $f$.
Let $a \neq \mathbb{I}$ be an element in $A$ with 
$\mu_f(a) = 1$ and $\nu(a) \leq 1$ for all $\nu \in S_A$.
Then $a = s_1 e + s_2 e'$ with an atom $e$, $s_1,s_2 \leq 1$
and $s_1 < 1$ or $s_2 < 1$.
We have to show the uniqueness of 
$\left\{\rho \in A^{*} : \rho(\mathbb{I})=1=\rho(a) \right\}$.

Suppose $s_1 < 1$. 
From $1 = s_1 \mu_f(e) + s_2 \mu_f(e')$ 
and $1 = \mu_f(e) + \mu_f(e')$
we get that $\mu_f(e) = 0$ and $\mu_f(e') = 1 = s_2$.
Hence $\mu_f = \mu_{e'}$.
Since the state space is strong, we have 
$f = e'$ and $a = s_1 f' + f$. 
In the case $s_1 = 0$ we get $a = f$.
In the case $s_1 \neq 0$ we get
for $\rho \in A^{*}$ with  $\rho(\mathbb{I}) = 1$ that
$1 = \rho(a)$ iff $\rho(f') = \rho(a) - \rho(f) = s_1 \rho(f')$ 
iff $0 = \rho(f') $ iff $ 1 = \rho(f)$.
Therefore
$\left\{\rho \in A^{*} : \rho(\mathbb{I})=1=\rho(a) \right\} 
= \left\{\rho \in A^{*} : \rho(\mathbb{I})=1=\rho(f) \right\} $
in all cases with  $s_1 < 1$.

The same follows in the case $s_2 < 1$ with exchanged roles
of $e$ and $e'$ and of $s_1$ and $s_2$. 

(iii) Assume that equation (6) holds
for all atoms $e$ and $f$. This means 
$\mu_e(f) + \mu_{e'}(f) = 1 = \mu_e(f') + \mu_{e'}(f')$.
Select any atom $e$ and consider 
$$X := \left\{ a \in A : \mu_e(a) + \mu_{e'}(a) = 0 \right\}.$$
Then $\mathbb{I} \notin X$ and $A = X \oplus \mathbb{R} \mathbb{I}$.

If $f$ is any atom, $f = \frac{1}{2}(x + \mathbb{I})$
with $x \in X$ and $\left\| x \right\| = 1$;
choose $x := f - f'$.
Vice versa, if $x \in X \subseteq A$ and $\left\| x \right\| = 1$, 
let $x = s f + t f'$ 
with an atom $f$ and $s,t \in \mathbb{R}$
be the spectral decomposition of $x$. With 
$\mu_e + \mu_{e'}$ we get $0 = s + t$. 
Therefore $ x = s (f - f')$ and
$ 1 = \left\|x\right\| = \left|s\right| $.
This means that either $s = 1$, $x = f - f'$ and $\frac{1}{2}(x + \mathbb{I}) = f$
or $s = -1$, $x = f' - f$ and $\frac{1}{2}(x + \mathbb{I}) = f'$.
Therefore, the atoms in $L_A$ are identical with the 
$\frac{1}{2}(x + \mathbb{I})$, $x \in X$, $\left\|x\right\| = 1$.
For $0 \neq x \in X$ and $s \in \mathbb{R}$ the
spectral decomposition of $ x + s $ is 
$$x + s \mathbb{I} = (s + \left\|x\right\|) f + (s - \left\|x\right\|)f'$$ 
with the atom $$f := \frac{1}{2}\left(\frac{1}{\left\|x\right\|} x + \mathbb{I}\right).$$
Then $0 \leq x + s \mathbb{I}$
iff  $0 \leq s - \left\|x\right\|$.
Use $\mu_{f'}$ for the only-if direction.

Assume $x \in X$, $\left\|x\right\| = 1$, and choose
an element $\rho$ in the dual of $X$ 
with $1 = \rho(x) = \left\|\rho\right\|$. 
Consider the atom
$f = \frac{1}{2}(x + \mathbb{I})$ 
and the following extension $\tilde{\rho}$ of $\rho$ to a state on $A$ and $L_X$;
$\tilde{\rho} : x + s \mathbb{I} \rightarrow \rho(x) + s$ 
for $x \in X$ and $s \in \mathbb{R}$.
Then $\tilde{\rho}(f) = 1$ and thus $\tilde{\rho} = \mu_f$.
Therefore $\tilde{\rho}$ and thus $\rho$ are uniquely determined 
and $X$ is smooth.
\newpage

Now assume $\left\|t x + (1-t) y\right\| = 1$ with 
$x,y \in X$, $\left\|x\right\| = 1 = \left\|y\right\|$ and $0 < t < 1$.
Select an element $\rho$ in the dual of $X$ 
with $\rho(t x + (1-t) y) = 1 = \left\|\rho\right\|$.
Then $1 = t \rho(x) + (1-t) \rho(y)$ with $ \left|\rho(x)\right| \leq 1$ 
and $ \left|\rho(y)\right| \leq 1$. Therefore $\rho(x) = 1 = \rho(y)$.
Use the same extension $\tilde{\rho}$ of $\rho$ to a state on $A$ as above
and the atoms $f_1 = \frac{1}{2}(x + \mathbb{I})$ and $f_2 = \frac{1}{2}(y + \mathbb{I})$.
Then $\tilde{\rho}(f_1) = 1 = \tilde{\rho}(f_2)$ and 
$\mu_{f_1} = \tilde{\rho} = \mu_{f_2}$.
Since $S_A$ is strong, we get
$f_1 = f_2$ and $x = y$.
We have thus shown that $X$ is strictly convex,
which completes the proof that $A$ is a generalized spin factor.

The only-if parts of (iii) and (iv) have been proven in section 7.
For the proof of the remaining part of (iv)
assume that the transition probabilities are symmetric.
For any two atoms $e$ and $f$ then
$\mathbb{P}(f|e) + \mathbb{P}(f|e') 
= \mathbb{P}(e|f) + \mathbb{P}(e'|f) 
= \mathbb{P}(\mathbb{I}|f)
= 1$.
This means that (6) holds.
By (iii) $A$ is a generalized spin factor, and 
by Theorem 7.1 $A$ is a spin factor.

(v) If $A$ is a spin factor, $A$ is a Jordan algebra 
and the product is bilinear.
Now suppose that the product is bilinear.
Choose any atom $f$ and define an inner product by
$\left\langle a|b\right\rangle := \mu_f(a \circ b) + \mu_{f'}(a \circ b)$
for $a,b \in A$. With $a = se + te'$, $s,t \in \mathbb{R}$ 
and an atom $e$ we then have 
$\left\langle a|a \right\rangle 
= s^{2} (\mu_f(e) + \mu_{f'}(e)) + t^{2} (\mu_f(e') + \mu_{f'}(e'))$.

Since $0 = \mu_f(e) + \mu_{f'}(e)$ would imply 
$\mu_f(e) = 0 = \mu_{f'}(e)$, $\mu_f(e') = 1 = \mu_{f'}(e')$
and thus $\mu_f = \mu_{e'} = \mu_{f'}$,
which contradicts $\mu_f \neq \mu_{f'}$,
we have $0 < \mu_f(e) + \mu_{f'}(e)$.
In the same way we get 
$0 < \mu_f(e') + \mu_{f'}(e')$.

Therefore 
$\left\langle a|a \right\rangle = 0$ iff $s = 0 = t$ iff $a = 0$.
Choose $X := \left\{ x \in A : \left\langle \mathbb{I} |x\right\rangle = 0 \right\} 
= \left\{ x \in A : \mu_f(x) + \mu_{f'}(x) = 0 \right\}$
to see that $A$ is a spin factor.
\hfill $\square$
\vspace{0,3 cm}

We shall now show that 
the reverse of Theorem 9.1 (ii) also holds and
that each strictly convex and smooth compact convex set
is the state space of some binary quantum logic.
\vspace{0,3 cm}

\textbf{Theorem 9.2}
\itshape
Let $K$ be a strictly convex and smooth compact convex set
in a locally convex space $V$. Then there is a complete order unit space
$A_K$ with quantum logic $L_K = ext[0,\mathbb{I}]$ 
and state space $S_K$ such that the conditions (a), (b), (c) 
and (d) of Theorem 9.1 are satisfied
and the state space $S_K$ is isomorphic to $K$.

Moreover, 
$A_K$ is a generalized spin factor iff $K$ is the unit ball in some Banach space,
and $A_K$ is a spin factor iff $K$ is the unit ball in some Hilbert space.
\normalfont
\vspace{0,3 cm}

Proof. Suppose that $K$ is a strictly convex and smooth compact convex set
in a locally convex space $V$. Then the topological boundary $\partial K$ of $K$
coincides with the extreme boundary $ext(K)$.
Let $A_K$ consist of the continuous affine real functions on $K$.
Equipped with the usual pointwise ordering, 
$A_K$ becomes a complete order unit space; we can choose 
the constant function that allocates $1$ to each element of $K$
as order unit $\mathbb{I}$. Besides $\mathbb{I}$ and $0$,
the extreme points of the unit interval are those functions $e_\omega$
that assume the value $1$ at one point $\omega$ of the boundary
and the value $0$ at the antipodal point $\omega'$ \cite{AS01, AS02}.
\newpage

For $ \omega \in K$ define the state $\delta_\omega$ by
$\delta_\omega (a) := a(\omega)$, $a \in A_K$.
The map $\omega \rightarrow \delta_\omega$ 
is an isomorphism between $K$ and the state space $S_K$ of $A_K$ \cite{AS02}
and $S_K$ is strong for the quantum logic $L_K$.
Each $a \in A_K$ assumes its maximum $s$ at a point $\omega$ on the boundary 
and its minimum $t$ at the antipodal point $\omega'$;
then $a = s e_\omega + t e_{\omega'}$. Moreover,
$\mathbb{P}(e_{\omega_2}|e_{\omega_1}) = \delta_{\omega_1} (e_{\omega_2})$
for $\omega_1,\omega_2 \in \partial K$.

Suppose that $K$ is the unit ball in some Banach space.
Then $-\omega_1$ is the antipodal point to a boundary point $\omega_1$ and 
$\frac{1}{2}(\delta_{\omega_1} + \delta_{-\omega_1}) = \delta_0$ on $A_K$.
For all boundary points $\omega_2$ we have
$\delta_0(\epsilon_{\omega_2}) = 1/2$ and
$\delta_{\omega_1}(\epsilon_{\omega_2}) + \delta_{-\omega_1}(\epsilon_{\omega_2}) 
= 2 \delta_0(\epsilon_{\omega_2}) = 1$.
Therefore equation (6) is satisfied and, by Theorem~9.1~(iii),
$A_K$ becomes a generalized spin factor $X \oplus \mathbb{R}$ 
with a smooth and strictly convex normed linear space $X$.
The unit ball of the dual $X^{*}$ is isomorphic to 
the state space of $A_K$ and thus to $K$.
If $K$ is the unit ball in a Hilbert space space,
$X^{*}$ becomes a Hilbert space. 
Then the second dual $X^{**}$ is a Hilbert space as well. 
The canonical embedding of $X$ in $X^{**}$ 
shows that $X$ is a pre-Hilbert space.
Moreover, since $A_K$ is complete, $X$ must be complete
and we get a Banach space $X$ in the first case here
and a Hilbert space $X$ in the second case.

If $A_K$ is a generalized spin factor, 
arising from the smooth and strictly convex normed linear space $X$,
its state space is isomorphic to the unit ball 
of the dual of $X$ and this dual is a Banach space.
If $A_K$ is a spin factor, $X$ is a pre-Hilbert space
and its dual is a Hilbert space.
\hfill $\square$
\vspace{0,3 cm}

With Theorem 9.1, we have not only characterized the generalized
spin factors in a new axiomatic way, but we have 
found a more general type of binary model.
Beyond that, we have identified
a one-to-one correspondence between the models of this type
and the strictly convex and smooth compact convex sets.
A model of this type arises from each such set $K$ by Theorem 9.2.
The characteristics of $K$ determine those of the transition probabilities;
they are symmetric iff $K$ is the unit ball in a Hilbert space,
and they satisfy equation~(6) for all atoms $e$ and $f$ iff $K$ 
is the unit ball of a Banach space.

The proof of Theorem~9.2 reveals that our transition probability
becomes identical with the one 
introduced by Mielnik \cite{mielnik1968geometry, mielnik1969theory}
in a different way. He considers a convex set and directly defines the transition probability between its 
extreme points (the pure states).

\section{Examples arising from the spaces $l^{p}$ and $L^{p}$}

By Theorem 7.1 non-symmetric transition probabilities
arise when we construct a generalized spin  factor 
from  a smooth and strictly convex normed space $X$ 
that is not a Hilbert space. Such spaces are 
the $L^{p}$ and $l^{p}$ spaces
with the norm $\left\| \ \right\|_p$,
$1 < p < \infty$ and $p \neq 2$.
With $p=2$, $L^{p}$ and $l^{p}$ become Hilbert spaces and
the transition probabilities are symmetric.
The dual spaces of $L^{p}$ and $l^{p}$ are $L^{q}$ and $l^{q}$
with $\frac{1}{p} + \frac{1}{q} = 1$. 
For $(\alpha_k) \in l^{p}$ and $(\beta_k) \in l^{q}$ we have the duality
\begin{center}
$\left\langle (\beta_k)| (\alpha_k) \right\rangle = \sum_k \alpha_k \beta_k$
\end{center}
\begin{center}
and $ \left\|(\alpha_k)\right\|^{p}_{p} = \left\langle (\beta_k)| (\alpha_k) \right\rangle$
with $(\beta_k) = (sign(\alpha_k) \left|\alpha_k\right|^{p-1}) \in l^{q}$.
\end{center}
Here, $sign(\alpha) := 1, 0, -1$ for the three cases $\alpha > 0, = 0, < 0$ 
with any real number~$\alpha$. 

For $x,y \in l^{p}$ with $\left\|x\right\|_p = \left\|y\right\|_p = 1$ and the 
corresponding atoms $e = \frac{1}{2}(x \oplus 1)$, $f = \frac{1}{2}(y \oplus 1)$
in the quantum logic $L_X$ with $X = l_p$,
we get by (5)
$$\mathbb{P}(f|e) = \frac{1}{2}\left(1 + \sum_k sign(\alpha_k) \left|\alpha_k\right|^{p-1} \beta_k \right)$$
and
$$\mathbb{P}(e|f) = \frac{1}{2}\left(1 + \sum_k \alpha_k sign(\beta_k) \left|\beta_k\right|^{p-1}\right),$$
when $x = (\alpha)_k$, $y = (\beta)_k$. 
In the case $\alpha_1 = 1$ and $\alpha_k = 0$ for $k \neq 1$ this becomes
\begin{center}
$\mathbb{P}(f|e) = \frac{1}{2}(1 + \beta_1) $ 
and 
$\mathbb{P}(e|f) = \frac{1}{2}(1 + sign(\beta_1) \left|\beta_1\right|^{p-1}) $.
\end{center}
\begin{figure}[ht]
\centering
\includegraphics[width=10cm]{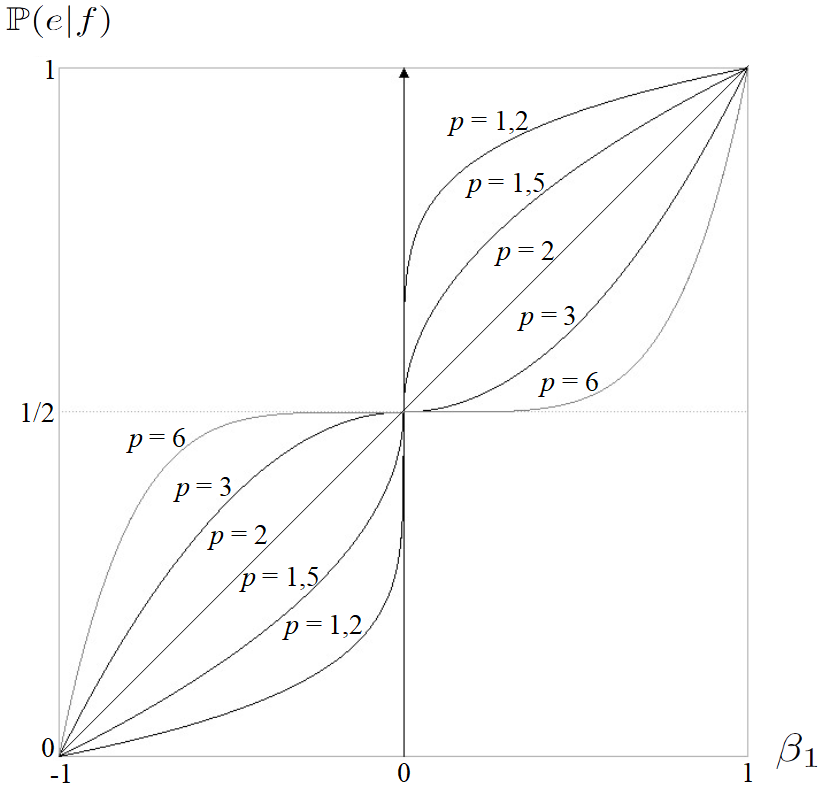}
\caption{Transition probability in the quantum logic $L_X$ with $X = l^{p}$}
\end{figure}

Figure 1 shows this transition probability $\mathbb{P}(e|f)$
as a function of the parameter $\beta_1$ for some different $p$.
The diagonal line represents not only the 
symmetric case $p = 2$ for $\mathbb{P}(e|f)$, 
but as well the transition probability $\mathbb{P}(f|e)$, which does not
depend on $p$ in the case considered here. The deviation of $\mathbb{P}(e|f)$
from $\mathbb{P}(f|e)$ increases, when $p$ moves away from $2$, 
either towards larger numbers or closer to 1.
The difference between $\mathbb{P}(e|f)$ and $\mathbb{P}(f|e)$ 
can come close to $\frac{1}{2}$,
when $p$ approaches $\infty$ or $1$.
In the limiting case $p \rightarrow \infty$, $\mathbb{P}(e|f)$ 
would become $\frac{1}{2}$ with the three exceptions $\beta_1 = -1, 0, 1$
where we always have $\mathbb{P}(e|f) = 0, \frac{1}{2}, 1$. 
Note that $\beta_1 =1$ means $e = f$ and that $\beta_1 = -1$
means $e' = f$.
In the other limiting case $p \rightarrow 1$, $\mathbb{P}(e|f)$ 
would become $0$ for $-1 < \beta_1 < 0$ and $1$ for  $0 < \beta_1 < 1$.

The striking symmetry of each curve with respect to the point $(0, 1/2)$
in Figure 1 is due to equation (6); a swap of $f$ and $f'$ means 
a change of sign for~$\beta_1$.

We complete this section with a brief look at the 
spaces $L^{p}$, $1 < p < \infty$, with any measure $\lambda$. 
In the same way as above we get the transition probabilities
$$\mathbb{P}(f|e) = \frac{1}{2}\left(1 + \int g \ sign(h) \left|h\right|^{p-1} d\lambda \right)$$
and
$$\mathbb{P}(e|f) = \frac{1}{2}\left(1 + \int sign(g) \left|g\right|^{p-1} h \ d\lambda \right)$$
for any $g,h \in L^{p}$ with $\left\|g\right\|_p = \left\|h\right\|_p = 1$ and the 
corresponding atoms $e = \frac{1}{2}(g \oplus 1)$, $f = \frac{1}{2}(h \oplus 1)$ 
in the quantum logic $ext[0,\mathbb{I}]$ of the generalized spin factor $A = L^{p} \oplus \mathbb{R}$.

The spaces $l^{1}$, $L^{1}$, $l^{\infty}$ and $L^{\infty}$ are 
neither smooth nor strictly convex and the 
transition probabilities do not exist for the atoms 
in the quantum logic $ext[0,\mathbb{I}]$ in
$A = l^{1} \oplus \mathbb{R}$, $A = L^{1} \oplus \mathbb{R}$, 
$A = l^{\infty} \oplus \mathbb{R}$ or $A = L^{\infty} \oplus \mathbb{R}$.

\section{Further examples}

Binary examples where (6) is violated can easily be constructed 
using Theorem 9.2 and a strictly convex and smooth compact convex set
$K$ that is not isomorphic to the unit ball of some normed space $X$.

A non-binary example with non-symmetric transition probability 
becomes Alfsen and Shultz's \cite{AS02} triangular pillow 
shown in Figure 2 
(more precisely: the quantum logic is not the pillow itself, 
but the orthomodular lattice consisting of the 
projective faces, the compressions or projective units; 
the triangular pillow is its state space).
We avoid to go into the details of Alfsen and Shultz's
theory and sketch this example only briefly.

The vertexes of the equilateral triangle represent three pairwise orthogonal
atoms $f_1,f_2,f_3$ with $ f_1 + f_2 + f_3 = \mathbb{I}$. A boundary
point on the curved surface off the triangle and its antipodal boundary
point represent two orthogonal atoms the sum of which is $\mathbb{I}$;
this means that the orthogonal complement of a boundary point 
on the curved surface off the triangle
is its antipodal point. An example of two such atoms are the north-pole point $e$ 
and the south-pole point $e'$.

The triangular pillow 
is not strictly convex at the edges 
and not smooth smooth at the vertexes of the triangle.
However, it is a \emph{spectral} convex set
and, for each atom, there is only one single state 
such that the atom carries the probability $1$ in this state~\cite{AS02}. 
Therefore, the transition probabilities exist
for the atoms. Lemma 3.2 then rules out that they
are symmetric, since we have $\mathbb{I} = f_1 + f_2 +f_3 = e + e'$
for the three pairwise orthogonal atoms $f_1, f_2, f_3$ and 
the orthogonal pair of atoms $e, e'$. 

Moreover, the symmetries of the triangular pillow imply
$\mathbb{P}(f_1|e) = \mathbb{P}(f_2|e) = \mathbb{P}(f_3|e)
= \mathbb{P}(f_1|e') = \mathbb{P}(f_2|e') = \mathbb{P}(f_3|e')$ and 
$\mathbb{P}(e|f_k) = \mathbb{P}(e'|f_k) $ for $k = 1,2,3$.
Since $\mathbb{P}(f_1|e) + \mathbb{P}(f_2|e) + \mathbb{P}(f_3|e) = \mathbb{P}(\mathbb{I}|e) = 1$
and $\mathbb{P}(e|f_k) + \mathbb{P}(e'|f_k) = \mathbb{P}(\mathbb{I}|e) = 1 $,
we get for $k = 1,2,3$ the non-symmetric transition probabilities
\begin{center}
$\mathbb{P}(f_k|e) = \mathbb{P}(f_k|e') =1/3$ 
and 
$\mathbb{P}(e|f_k) = \mathbb{P}(e'|f_k) = 1/2$ 
\end{center}
and furthermore 
\begin{center}
$\sum_k \mathbb{P}(e|f_k) = \sum_k \mathbb{P}(e'|f_k) = 3/2$
and $\mathbb{P}(f_k|e) + \mathbb{P}(f_k|e') = 2/3$,
\end{center}
which violates equation (6) in section 7.

The triangular pillow does not satisfy 
the conditions (b) and (c) of Theorem~9.1,
since $f_1 + f_2$ is not an atom, but 
$0 \neq f_1 + f_2 \neq 1$ and $f_2 \neq {f_1}'$. It
is one very demonstrative three-dimensional example
in a larger class which Alfsen and Shultz constructed \cite{AS02}
and which includes spectral convex sets with higher dimensions.

\begin{figure}[ht]
\centering
\includegraphics[width=6.5cm]{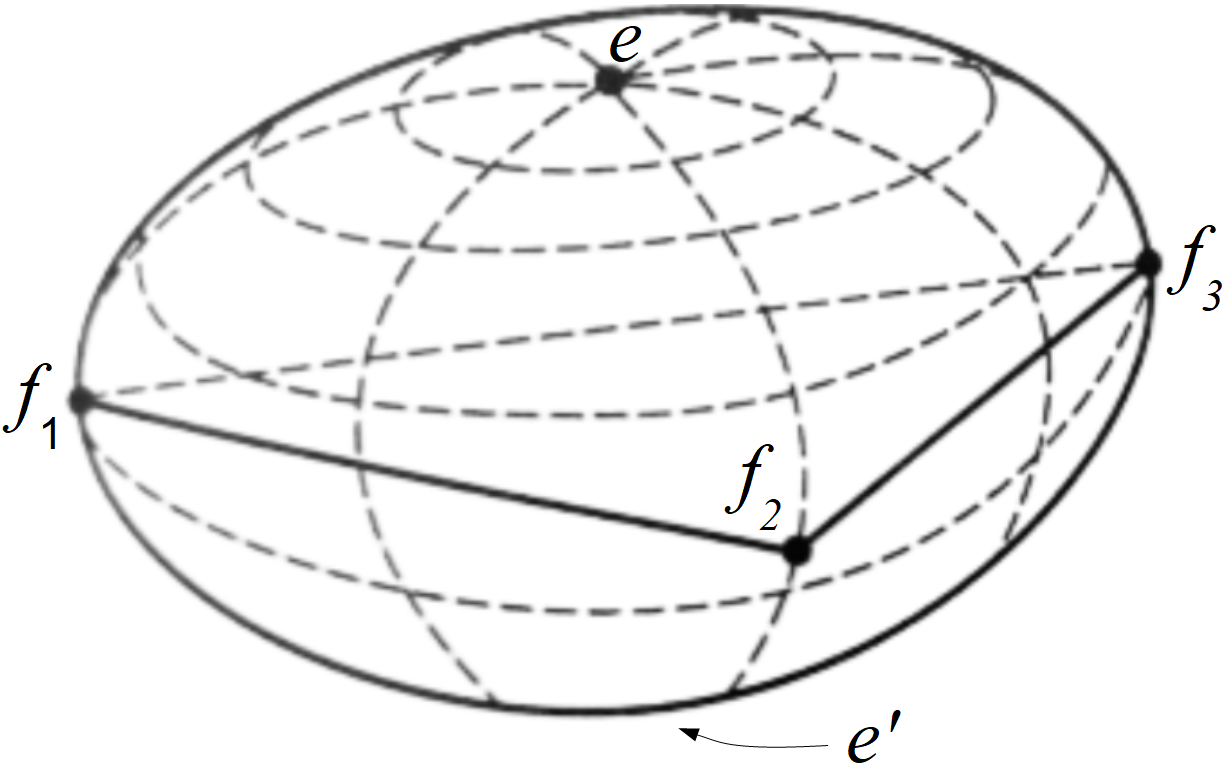}
\caption{The triangular pillow \cite{AS02}}
\end{figure}

The examples of this section yield further cases, 
where Segal's postulates~\cite{segal1947postulates} are satisfied, 
but a bilinear product that is in line with the 
spectral decomposition does not exist (see section 8).
This follows for the first examples 
from Theorem 9.1 (v) and for the triangular pillow from 
Corollary 9.44 in Ref.~\cite{AS02}.
The product is bilinear then only for 
compatible elements of the order unit space;
these are linear combinations of the same pairwise orthogonal 
elements $e_k$ of the quantum logic and then
$\left( \sum s_k e_k \right) \circ \left( \sum t_k e_k \right) = \sum s_k t_k e_k$ 
for $s_k,t_k \in \mathbb{R}$.
A bilinear product for elements that are not compatible
may be very appealing from the mathematical point of view,
but convincing physical reasons 
to make it a general postulate 
are hard to find.
\newpage
\section{Conclusions}

We have seen that there is an abundance of mathematical structures 
with non-symmetric transition probabilities, and our main result is 
a complete axiomatic characterization of the binary models,
which represent the classical bit, the quantum bit (qubit) 
and generalized versions thereof.
Their state spaces are smooth and strictly 
convex compact convex sets and, vice versa, 
for each such set $K$ there is a model with state space $K$.
This reveals an interesting bidirectional relation 
between the binary models and 
the smooth strictly convex compact convex sets.
The characteristics of $K$ determine those of the transition probabilities.
They satisfy equation (6) for all atoms $e$ and $f$ if and only if 
$K$ is the unit ball of a Banach space,
and they become symmetric if and only if 
$K$ is the unit ball of a Hilbert space.
The first case results in a generalized spin factor 
and the second case in a genuine spin factor.

Only the genuine spin factors possess a bilinear product
and become formally real Jordan algebras.
They include the real version of the qubit, the usual complex qubit 
and many further cases. Some favorable features 
distinguishing the usual complex qubit
(and, more generally, quantum mechanics with the Hilbert spaces over the complex numbers)
have been identified: the generator of every continuous reversible time evolution 
can be associated to an observable~\cite{barnum2014higher}, and
the usual qubit permits a 
reasonable model of a multiple bit system 
with local accessibility of the state \cite{wootters1986quantum, wootters1990local};
nowadays local accessibility has become more familiar 
as local tomo\-graphy~\cite{barnum2014higher}. 
Nevertheless the other spin factors play an important role
in quantum mechanics, since they represent the canonical anti-commutator relations for
the fermions. 

It would be interesting to know
whether there are any further quantum logics 
with strong state spaces and 
non-symmetric transition probabilities 
beyond those considered here
(and their direct sums), 
perhaps with a richer structure 
and an infinite family of 
pairwise orthogonal atoms.
A next step might be to study generalized ternary models
along the lines of Theorem~9.1, which 
means that the assumptions (b) and (c) in Theorem~9.1 
are replaced by: 
\begin{enumerate}
\item[(b')]
\itshape
The maximum number of pairwise orthogonal atoms is three.
\normalfont
\item[(c')]
\itshape
Each $a \in A$ has a spectral decomposition
$a = s_1 e_1 + s_2 e_2 + s_3 e_3 +$ with $s_1, s_2, s_3 \in \mathbb{R}$ 
and three pairwise orthogonal atoms $e_1, e_2, e_3$.
\end{enumerate}
\normalfont
This includes the decomposable cases 
$\mathbb{R} \oplus \mathbb{R} \oplus \mathbb{R}$, 
$A \oplus \mathbb{R}$,
where $A$ is an order unit space satisfying
the assumptions (a), (b), (c) and (d) in Theorem 9.1,
and the four non-decomposable formally real Jordan algebras
formed by the self-adjoint $3 \times 3$-matrices over 
the real numbers, complex number, quaternions or 
octonions~\cite{hanche1984jordan, nie2021generic}.
The first case is the classical one, the complex
$3 \times 3$-matrices
represent the usual quantum mechanical ternary model
(the complex qutrit).
In the four Jordan algebras,
the transition probability is symmetric~\cite{nie2021generic}. 
An open issue is whether these are already all or 
whether there are more non-decomposable ternary cases,
possibly with non-symmetric transition probabilities, 
and whether they can be classified.
Note that the triangular pillow
does not satisfy (c'); it represents neither a generalized binary
nor ternary model, but a strange non-decomposable hybrid between these two.
Alfsen and Shultz's further examples in the class where the 
triangular pillow belongs to \cite{AS02}
are hybrids with higher dimensions.
\vspace{0,6 cm}

\noindent
\textbf{STATEMENTS AND DECLARATIONS}
\vspace{0,2 cm}

\noindent
\textbf{Data accessibility.} This article has no additional data.

\noindent
\textbf{Competing interests.} The author declares he has no competing interests.

\noindent
\textbf{Funding.} No funding has been received for this article.

\bibliographystyle{abbrv}
\bibliography{Literatur2022}
\end{document}